\documentclass[twocolumn]{aastex701}
\usepackage{graphicx,amsmath,amsfonts,amssymb,multirow,mathrsfs,ulem}
\usepackage{hyperref}
\usepackage{epstopdf}

\begin{document}

\title{Joint Constraints on Fuzzy and Warm Dark Matter from Satellite Populations of \\the Milky Way and Andromeda}

\correspondingauthor{Yan Gong}
\email{Email: gongyan@bao.ac.cn}

\author[0009-0004-4092-3931]{Jianxiang Liu}
\affiliation{National Astronomical Observatories, Chinese Academy of Sciences, 20A Datun Road, Beijing 100101, China}
\affiliation{School of Astronomy and Space Sciences, University of Chinese Academy of Sciences, \\No.19A Yuquan Road, Beijing 100049, China}
\email{liujx@bao.ac.cn}

\author[0000-0003-0709-0101]{Yan Gong}
\affiliation{National Astronomical Observatories, Chinese Academy of Sciences, 20A Datun Road, Beijing 100101, China}
\affiliation{School of Astronomy and Space Sciences, University of Chinese Academy of Sciences, \\No.19A Yuquan Road, Beijing 100049, China}
\affiliation{Science Center for Chinese Space Station Survey Telescope, National Astronomical Observatories, \\Chinese Academy of Science, 20A Datun Road, Beijing 100101, China}
\email{gongyan@bao.ac.cn}

\author[0000-0002-4359-5994]{Kai Liao}
\affiliation{School of Physics and Technology, Wuhan University, Wuhan 430072, China}
\email{liaokai@whu.edu.cn}

\begin{abstract}
We perform a joint analysis of the Milky Way (MW) and Andromeda (M31) satellite populations to constrain the properties of fuzzy dark matter (FDM) and thermal-relic warm dark matter (WDM). 
We combine MW satellite observations from the Dark Energy Survey (DES) and Pan-STARRS1 (PS1) with M31 satellite data from the Pan-Andromeda Archaeological Survey (PAndAS), and model the corresponding observable satellite populations using the empirical galaxy--halo connection model described in Nadler et al. (2020) together with the appropriate selection functions. 
Uncertainties in the virial masses of the MW and M31 are incorporated through host-mass priors that linearly scale the relevant model parameters, allowing us to infer the full posterior distributions of all parameters.
For the FDM case, we obtain $m_{\mathrm{FDM}} > 1.74 \times 10^{-20}~\mathrm{eV}$ 
(95\% CL) and $m_{\mathrm{FDM}} > 1.42 \times 10^{-20}~\mathrm{eV}$ 
(20:1 posterior ratio). For thermal-relic WDM, we find 
$m_{\mathrm{WDM}} > 6.20~\mathrm{keV}$ (95\% CL) and 
$m_{\mathrm{WDM}} > 5.75~\mathrm{keV}$ (20:1 posterior ratio).
These results represent a moderate improvement over MW-only constraints, and provide the strongest constraints to date on the FDM and WDM derived from satellite galaxy populations in the Local Group. 

\end{abstract}


\section{Introduction} \label{sec1}
Dark matter (DM) constitutes the dominant component of matter in the Universe \citep{2020A&A...641A...6P}. 
Its existence is supported by multiple cosmological and astrophysical observations, such as the cosmic microwave background (CMB; \citealt{2020A&A...641A...1P}) and galaxy rotation curves \citep{1970ApJ...159..379R,1985ApJ...295..305V,2015PASJ...67...75S}, yet its true nature remains elusive. 
The cold dark matter (CDM) paradigm, exemplified by weakly interacting massive particles (WIMPs), has achieved remarkable success in explaining the large-scale structure (LSS) of the Universe \citep{1996PhR...267..195J,2005Natur.435..629S,2012AnP...524..507F,2014MNRAS.444.1518V,2018MNRAS.475..624N}.
However, it faces several well-known small-scale challenges: the missing-satellite problem, the cusp--core problem, and the too-big-to-fail problem \citep{2017Galax...5...17D,2017ARA&A..55..343B,2026NatAs.tmp....5V}.

To address these issues and motivated by the lack of any experimental detection of CDM particles, several non-CDM models have been proposed or revisited, such as fuzzy dark matter (FDM; \citealt{2000PhRvL..85.1158H,2021ARA&A..59..247H,2025arXiv250700705E}), warm dark matter (WDM; \citealt{2019PrPNP.104....1B}), and interacting dark matter (IDM; \citealt{2001PhLB..518....8B, 2018PhRvL.121h1301G}). 
These non-CDM scenarios reproduce the LSS of the Universe while simultaneously altering the linear matter power spectrum and the abundance of halos and subhalos on small scales.
We should note that the role of baryonic physics is also significant at small scales, which can alleviate some of the small-scale challenges faced by the CDM model without the necessity of invoking non-CDM \citep{2014MNRAS.437..415D,2015MNRAS.454.2981C,2017ARA&A..55..343B,2022NatAs...6..897S}. Therefore, the main reason for these issues remains a subject of ongoing debate and is yet to be fully resolved \citep{2017ARA&A..55..343B,2022NatAs...6..897S}.

Dwarf galaxies, as low-mass and DM-dominated systems, serve as powerful probes of small-scale dark matter structure \citep{2019PhRvL.123e1103M,2021PhRvL.126i1101N,2025PhRvL.134o1001Z}. 
Following the success of the Sloan Digital Sky Survey (SDSS) in the early 2000s \citep{2005AJ....129.2692W,2005ApJ...626L..85W}, an increasing number of dwarf galaxies, including progressively fainter systems, have been discovered and continue to be found \citep{2009Natur.461...66M,2016MNRAS.460.1270D,2016arXiv161205560C,2018ApJ...868...55M,2020ApJ...893...47D,2022ApJ...933..135D,2023MNRAS.523..876Q,2025OJAp....8E..89T,2025arXiv250704618C,2025arXiv250912313T}, further motivating the use of dwarf galaxies to study dark matter models. 
Non-CDM models can suppress the abundance of halos and subhalos, and since satellite dwarf galaxies reside within subhalos, it is well-motivated to use their observed abundance to probe the underlying subhalo population, thereby placing constraints on non-CDM models \citep{2021PhRvL.126i1101N,2021JCAP...08..062N,2022PhRvD.106l3026D,2025PhRvD.111f3079T,2025ApJ...986..127N}.

For instance, \citet{2021PhRvL.126i1101N} and \citet{2025PhRvD.111f3079T} use the empirical galaxy--halo connection model  introduced in \citet{2020ApJ...893...48N} together with observations of Milky Way (MW) satellites from the Dark Energy Survey (DES; \citealt{2016MNRAS.460.1270D}) and Pan-STARRS1 (PS1; \citealt{2016arXiv161205560C}) to constrain various non-CDM models. \citet{2025ApJ...986..127N} use the same modeling framework and data, incorporating updated transfer functions and subhalo mass functions to constrain FDM, WDM, and IDM. In addition, \citet{2021JCAP...08..062N} and \citet{2022PhRvD.106l3026D} develop different semi-analytic frameworks and use the predicted total number of MW satellites from \citet{2018MNRAS.479.2853N} and \citet{2020ApJ...893...48N} to place constraints on WDM.

In this paper, we present the joint constraints on non-CDM models using the satellite populations of the MW and Andromeda Galaxy (M31).
Specifically, we adopt and slightly modify the empirical galaxy--halo connection model introduced by \citet{2020ApJ...893...48N} and use observations of MW satellites from DES and PS1 together with M31 satellites from the Pan-Andromeda Archaeological Survey (PAndAS; \citealt{2009Natur.461...66M,2018ApJ...868...55M,2022ApJ...933..135D,2023ApJ...952...72D}) to constrain the properties of FDM and thermal-relic WDM.
Using both satellite populations allows us to improve the overall precision of the constraints, compared to using only the MW satellite population \citep{2024ApJ...967...61N}.

This paper is organized as follows: 
in~Section~\ref{sec2}, we introduce the FDM and WDM models considered in this work; 
the observational data of satellite galaxies are presented in Section~\ref{sec3.2};
Section~\ref{sec3} describes the modeling framework; 
in~Section~\ref{sec4}, we show the constraint results on the FDM and WDM models;  
in~Section~\ref{sec5}, we discuss additional results derived from our modeling framework; 
we provide a summary and outlook in~Section~\ref{sec6}. 
Following \citet{2015ApJ...810...21M} and \citet{2025ApJ...986..127N}, we adopt the cosmological parameters $h = 0.7$, $\Omega_{\mathrm{m}} = 0.286$, $\Omega_{\mathrm{b}} = 0.049$, $\Omega_{\Lambda} = 0.714$, $\sigma_{8} = 0.82$, and $n_{\mathrm{s}} = 0.96$ \citep{2013ApJS..208...19H}.
This choice has a negligible impact on our results.
Halo masses are defined using the virial overdensity criterion of \citet{1998ApJ...495...80B}, which corresponds to $\Delta_{\mathrm{vir}} \approx 99\,\rho_{\mathrm{crit}}$ in our cosmology, where $\rho_{\mathrm{crit}}$ is the critical density of the Universe at $z = 0$.

\begin{figure*}
	\centering
	\begin{minipage}{\linewidth}
		\centering
		\includegraphics[width=\linewidth]{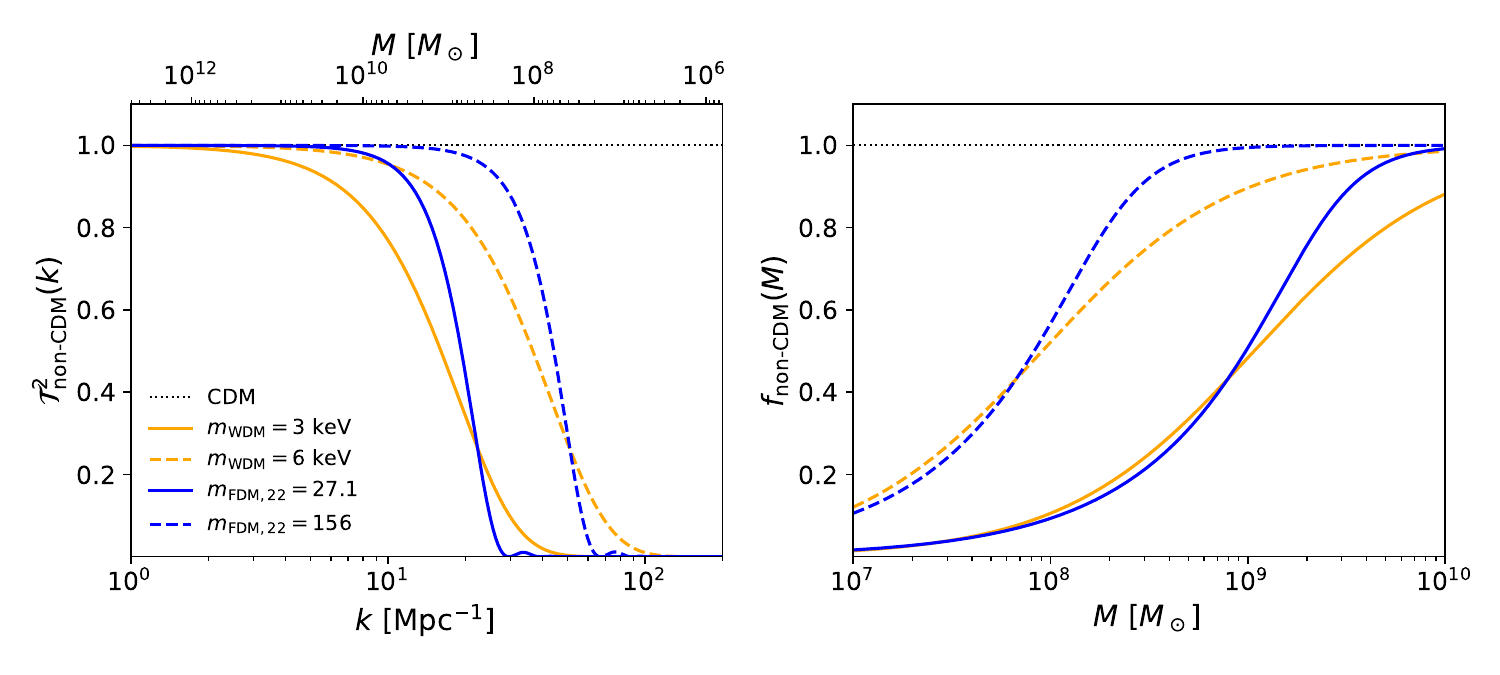}	
	\end{minipage}
    \caption{The left and right panels show the transfer functions and subhalo mass function suppressions for FDM (blue) and WDM (orange), respectively. In the left panel, the lower $x$-axis indicates the cosmological wavenumber, while the upper $x$-axis shows the corresponding halo mass. In the right panel, the $x$-axis represents the peak virial mass. Different line styles correspond to different particle masses, and models with the same line style have the same half-mode mass $M_{\mathrm{hm}}$. The solid and dashed curves correspond to $M_{\mathrm{hm}} = 4.30\times10^{8}~M_\odot$ and $3.64\times10^{7}~M_\odot$, respectively.}
    \label{transfer_shmf_relative.pdf}
\end{figure*}

\section{Dark Matter Models} \label{sec2}
In this section, we discuss the transfer functions and subhalo mass functions of the non-CDM models, and present the details of the FDM and WDM models we adopt in this work. 

\subsection{Non-cold dark matter models} \label{sec2.1}

For many non-CDM models, the linear matter power spectrum is identical to that of CDM on large scales but deviates on small scales. 
The FDM and WDM models considered in this work both can suppress small-scale power, through different physical mechanisms. 
To quantify their deviations from CDM, we normalize the linear matter power spectrum of each non-CDM model by that of CDM, defining the transfer function as
\begin{equation}
\mathcal{T}^2_\mathrm{non\text{-}CDM}(k) = \frac{P_\mathrm{non\text{-}CDM}(k)}{P_\mathrm{CDM}(k)},
\label{trans_func}
\end{equation}
where $k$ is the cosmological wave number, $P_\mathrm{CDM}(k)$ is the CDM linear matter power spectrum, and $P_\mathrm{non\text{-}CDM}(k)$ is the linear matter power spectrum of a non-CDM model \citep{2001ApJ...556...93B}.

The transfer functions we consider represent the suppression of the linear matter power spectrum in non-CDM models relative to CDM. 
To quantify the strength of this suppression, we introduce the half-mode wavenumber $k_\mathrm{hm}$, defined as the scale at which the power is reduced to one quarter of that in CDM \citep{2001ApJ...556...93B,2012MNRAS.424..684S}, which is given by
\begin{equation}
\mathcal{T}^2_\mathrm{non\text{-}CDM}(k_\mathrm{hm}) = 0.25.
\label{k_hm}
\end{equation}
In the linear theory, the halo mass and comoving wavenumber are related by \citep{2001ApJ...556...93B,2012MNRAS.424..684S,2019ApJ...878L..32N}
\begin{equation}
M(k) \equiv \frac{4\pi}{3} \Omega_{\mathrm{m}} \rho_{\mathrm{crit}} \left( \frac{\pi}{k} \right)^3.
\label{m_hm}
\end{equation}
The half-mode mass corresponding to $k_{\mathrm{hm}}$ is then defined as $M_{\mathrm{hm}} \equiv M(k_{\mathrm{hm}})$.

The non-CDM models suppress the number of subhalos by reducing the linear matter power spectrum on small scales, and the suppression of the subhalo mass function in non-CDM models relative to CDM, $f_{\mathrm{non\text{-}CDM}}(M)$, can be expressed as
\begin{equation}
\left( \frac{dN_{\text{sub}}}{dM} \right)_{\mathrm{non\text{-}CDM}} = f_{\mathrm{non\text{-}CDM}}(M) \left( \frac{dN_{\text{sub}}}{dM} \right)_{\text{CDM}},
\label{shmf}
\end{equation}
where $M$ is the peak virial mass \citep{2025ApJ...986..127N}. 
It should be noted that the specific forms of $\mathcal{T}^{2}_{\mathrm{non\text{-}CDM}}(k)$ and $f_{\mathrm{non\text{-}CDM}}(M)$ depend on the particular dark matter model and its underlying properties. 
Furthermore, for a specific non-CDM model, the half-mode mass $M_{\mathrm{hm}}$ derived from $\mathcal{T}^{2}_{\mathrm{non\text{-}CDM}}(k)$ provides a useful characteristic scale that facilitates the construction of $f_{\mathrm{non\text{-}CDM}}(M)$.
However, using $M_{\mathrm{hm}}$ for a direct comparison among different non-CDM models can be an oversimplification and may introduce significant biases in the resulting constraints \citep{2017JCAP...11..046M,2018PhRvD..98h3540M,2022JCAP...10..032H}.

\subsection{Fuzzy dark matter} \label{sec2.2}
FDM consists of ultra-light bosons with mass $m_\mathrm{FDM}\sim 10^{-22}~\mathrm{eV}$, whose de~Broglie wavelength is on the kiloparsec scale, comparable to the size of dwarf galaxies. 
Due to quantum pressure, the formation of small-scale structures is suppressed \citep{2000PhRvL..85.1158H,2021ARA&A..59..247H}. 
Following \citet{2025ApJ...986..127N}, we define
\begin{equation}
m_\mathrm{FDM,22} \equiv \frac{m_\mathrm{FDM}}{10^{-22}~\mathrm{eV}}.
\label{fdmmass}
\end{equation}
We adopt the FDM transfer function \citep{2022PhRvD.105l3529P}
\begin{equation}
\mathcal{T}^2_{\mathrm{FDM}}(k, m_{\mathrm{FDM,22}}) = \left[ \frac{\sin(x^m)}{x^m(1 + Bx^{6-m})} \right]^2,
\label{fdmtrans}
\end{equation}
where $m = 5/2$, $x \equiv A(k/k_\mathrm{J})$, $k_\mathrm{J} \equiv 9~\mathrm{Mpc}^{-1}\times m_\mathrm{FDM,22}^{1/2}$, $A=2.22(m_\mathrm{FDM,22})^{1/25-\ln(m_\mathrm{FDM,22})/1000}$, and $B=0.16(m_\mathrm{FDM,22})^{-1/20}$. 
We then obtain the half-mode mass, and it can be expressed as
\begin{equation}
M_\mathrm{hm}(m_\mathrm{FDM,22})=4.5\times10^{10}~M_\odot \times m_\mathrm{FDM,22}^{-1.41}.
\label{fdmmhm}
\end{equation}
Using the results from \citet{2025ApJ...986..127N}, the suppression of the subhalo mass function in FDM relative to CDM can be written as
\begin{equation}
f_{\mathrm{FDM}}(M,m_\mathrm{FDM,22}) = \left [ 1+\left ( \frac{\alpha M_\mathrm{hm}(m_\mathrm{FDM,22})}{M}  \right )^\beta  \right ]   ^{-\gamma},
\label{fdmshmf}
\end{equation}
where $\alpha = 5.5$, $\beta=2.5$ and $\gamma = 0.3$.

\subsection{Warm dark matter} \label{sec2.3}
In this work, we consider the simplest thermal-relic WDM, whose properties are determined solely by the particle mass $m_{\mathrm{WDM}}$, which is typically a few~keV.
The non-negligible free-streaming effect of WDM can suppress the formation of small-scale structures \citep{2019PrPNP.104....1B}.
We should notice that, although thermal-relic WDM is frequently employed as a benchmark for non-CDM models, the functional forms of the transfer functions for other scenarios may differ from the thermal-relic case \citep[see e.g.][]{2022PhRvL.129s1301Z}.

Following the same procedure as for FDM, we adopt the WDM transfer function from \citet{2005PhRvD..71f3534V}, which is given by
\begin{equation}
\mathcal{T}^2_{\mathrm{WDM}}(k, m_{\mathrm{WDM}}) = \left[1 + \left( \alpha(m_{\mathrm{WDM}}) \times k \right)^{2\nu} \right]^{-10/\nu},
\label{wdmtrans}
\end{equation}
where we take $\nu = 1.049$, and $\alpha(m_{\mathrm{WDM}})$ can be expressed as \citep{2023PhRvD.108d3520V}
\begin{equation}
\alpha(m_{\mathrm{WDM}}) = a \left( \frac{m_{\mathrm{WDM}}}{1 \,\mathrm{keV}} \right)^b \left( \frac{\omega_{\mathrm{WDM}}}{0.12} \right)^{\eta} \left( \frac{h}{0.6736} \right)^{\theta} ~ h^{-1} \, \mathrm{Mpc},
\label{wdmtrans2}
\end{equation}
where $a = 0.0437$, $b = -1.188$, $\eta = 0.2463$, $\theta = 2.012$, and $\omega_{\mathrm{WDM}} \equiv \Omega_{\mathrm{WDM}} h^2$. 
Then we can get the half-mode mass
\begin{equation}
M_{\mathrm{hm}}(m_{\mathrm{WDM}}) = 4.3 \times 10^8 ~ M_\odot \times \left( \frac{m_{\mathrm{WDM}}}{3 ~ \mathrm{keV}} \right)^{-3.564}.
\label{wdmmhm}
\end{equation}
Similarly, using the results from \citet{2025ApJ...986..127N}, the suppression of the subhalo mass function in WDM relative to CDM can also be expressed as
\begin{equation}
f_{\mathrm{WDM}}(M,m_\mathrm{WDM}) = \left [ 1+\left ( \frac{\alpha M_\mathrm{hm}(m_\mathrm{WDM})}{M}  \right )^\beta  \right ]   ^{-\gamma},
\label{wdmshmf}
\end{equation}
where $\alpha = 2.5$, $\beta=0.9$ and $\gamma = 1.0$.

It is worth noting that when characterizing the suppression of the subhalo mass function in FDM and WDM relative to CDM, \citet{2025ApJ...986..127N} utilized cosmological zoom-in simulations with modified initial conditions corresponding to FDM or WDM transfer functions. 
Although these simulations neglect FDM wave interference effects and the thermal velocities of WDM particles, this approximation is sufficiently accurate for the subhalos of interest in this work, which have peak virial masses of $M \gtrsim 10^8~M_{\odot}$ \citep{2017JCAP...11..017L,2020ApJ...893...48N,2023MNRAS.524.4256M}. 
The probabilistic inference was performed to reconstruct the model parameters ($\alpha$, $\beta$, and $\gamma$) which fit the simulated data, and the values that maximize the posterior distributions were adopted. 
They verified that the uncertainties associated with these parameters do not significantly affect the final constraints.
In~Figure~\ref{transfer_shmf_relative.pdf}, we show the examples of the transfer functions and subhalo mass function suppressions for the FDM and WDM models with the same $M_{\mathrm{hm}}$.
We can see that, for fixed $M_{\mathrm{hm}}$, the WDM exhibits a stronger suppression than the FDM at relatively high subhalo masses, whereas at lower masses its suppression becomes weaker than that of FDM.

\begin{figure}
	\centering
	\begin{minipage}{\linewidth}
		\centering
		\includegraphics[width=\linewidth]{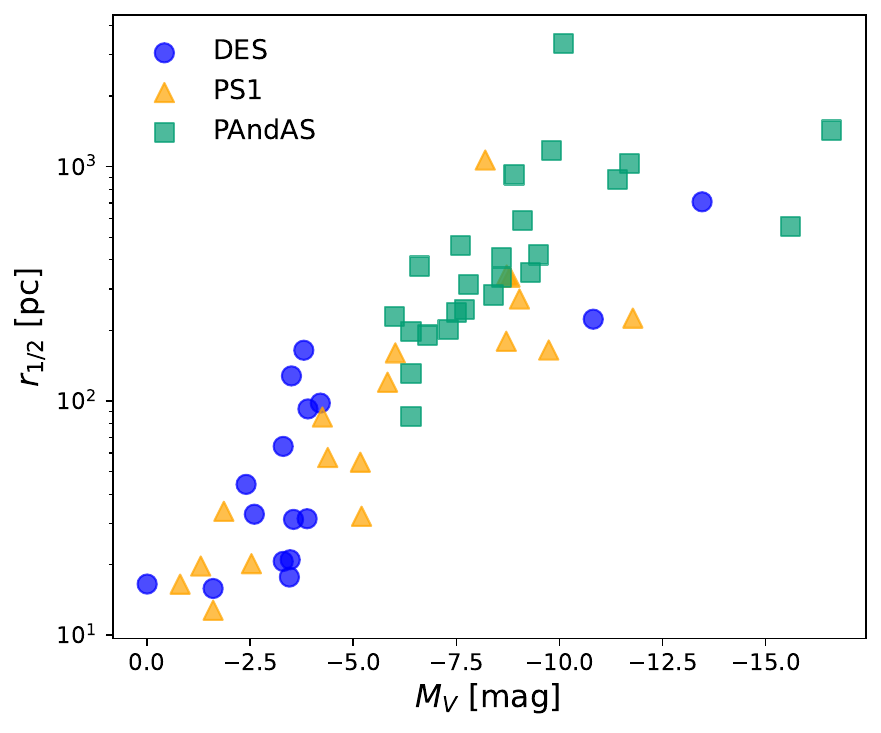}	
	\end{minipage}
    \caption{The half-light radii $r_{1/2}$ of the satellite galaxy data we use as a function of the absolute magnitudes $M_V$. The blue dots, orange triangles, and green squares mark the satellites from the DES, PS1, and PAndAS, respectively.}
    \label{satellite.pdf}
\end{figure}

\section{Observational data} \label{sec3.2}
We consider the observational data of satellite populations in both MW and M31 in the analysis. 
The observed MW satellite population used in this work is identical to that adopted in \citet{2020ApJ...893...48N}. 
It includes 18 satellites discovered in the PS1 DR1 data \citep{2012ApJ...750...99T} and 16 satellites identified in the DES Y3A2 data \citep{2018ApJS..239...18A}, yielding a total of 34 MW satellites. 
The faintest satellite in this sample, Cetus~II, has absolute magnitude $M_V \simeq 0~ \mathrm{mag}$ \citep{2015ApJ...813..109D}.

We use the DES and PS1 survey selection functions derived in \citet{2020ApJ...893...47D}, which are publicly available as machine-learning classifiers.\footnote{\url{https://github.com/des-science/mw-sats}} 
These classifiers predict the detection probability of a satellite as a function of its absolute magnitude $M_V$, heliocentric distance $D$, azimuthally averaged projected half-light radius $r_{1/2}$, and sky position. 
In addition, geometric constraints are applied to ensure that observable satellites lie within the survey footprints and to exclude regions where satellite detection is hindered by interstellar extinction, bright nearby stars, or bright extragalactic sources.
We only consider satellites located within $300~\mathrm{kpc}$ from the center of the MW.
In addition, to exclude likely star clusters, we restrict our sample to satellites with $r_{1/2}>10~\mathrm{pc}$ \citep{2020ApJ...893...48N}.

The observed M31 satellite population used in this work is drawn from the PAndAS survey data \citep{2023ApJ...952...72D}. 
We adopt the analytic observational selection function developed by \citet{2022ApJ...933..135D}.\footnote{\url{https://github.com/dolivadolinsky/Recovery_dwarf_galaxy_M31}} 
Although this selection function is analytic rather than a machine-learning classifier, it accounts for the dependence on absolute magnitude $M_V$, heliocentric distance $D$, azimuthally averaged projected half-light radius $r_{1/2}$, and sky position, with additional geometric constraints applied. 

Thus, although the methods for deriving the MW and M31 selection functions differ, both frameworks are designed to accurately capture the detection probability as a function of the key satellite properties. 
Moreover, the joint constraints are not dominated by the M31 selection function.
We therefore do not expect this methodological difference to introduce significant systematic bias into the joint inference of our model parameters. 
We have also verified through sensitivity test that the impact of this methodological difference is minimal.

Specifically, our M31 satellite sample consists of the full set of 24 satellites used in \citet{2023ApJ...952...72D}. 
The faintest satellite in our M31 sample is And~XXVI, with an absolute magnitude of $M_V \simeq -6.0~\mathrm{mag}$ \citep{2022ApJ...938..101S}. 
Although And~XXVI has a very low recovery fraction based on the selection function given in  \citet{2022ApJ...933..135D}, we have verified that including or excluding this system has a negligible impact on our results.
Due to the PAndAS survey footprint, we only consider satellites located between $30$ and $300~\mathrm{kpc}$ from the center of M31. 
Because of the limitations of the selection function, we additionally restrict our analysis to satellites with $r_{1/2}>63~\mathrm{pc}$, although we have confirmed that our results are insensitive to this cut.
In Figure~\ref{satellite.pdf}, we show the half-light radii $r_{1/2}$ as a function of the absolute magnitudes $M_V$ of all 58 satellite galaxies from DES, PS1, and PAndAS used in this work.

\section{Modeling Framework} \label{sec3}
In this section, we introduce our modeling framework based on the empirical galaxy--halo connection model \citep{2020ApJ...893...48N}, including the adopted dark matter-only zoom-in simulations, model parameters, and statistical inference procedure.\footnote{\url{https://github.com/eonadler/subhalo_satellite_connection}}  
More detailed description of the empirical galaxy--halo connection model can be found in \citet{2019ApJ...873...34N} and \citet{2020ApJ...893...48N}.

\subsection{Overview of the modeling framework} \label{sec3.1}
Our goal is to constrain dark matter models using the observed populations of satellite galaxies. 
To achieve this, we first construct a model capable of predicting satellite populations to fit the observational data. 
In this work, we use high-resolution dark matter-only zoom-in simulations based on CDM to generate subhalo populations. 
We then apply the empirical galaxy--halo connection model to populate these subhalos with satellite galaxies. 
The impact of different dark matter models is introduced through the suppression of the subhalo mass function in non-CDM model relative to CDM. 
By varying the model parameters, we can thus generate predicted satellite populations. 

After obtaining the predicted satellite populations, we convolve them with the observational selection functions, which quantify the probability of detecting satellites with different properties. 
This yields the predicted observable satellite populations.
With these predictions, we employ a Poisson-based likelihood function combined with suitable priors to derive the posterior distributions of the model parameters.

It is important to note that our approach assumes that the radial distribution of satellites remains unchanged across different dark matter models.
This assumption is justified for WDM \citep{2014MNRAS.439..300L,2017MNRAS.464.4520B}, while it is more uncertain for FDM. 
However, since the mean mass of the faintest satellites used in our analysis exceeds $10^{8}~M_{\odot}$ \citep{2020ApJ...893...48N}, the wave interference effects of FDM can be safely neglected within the mass range of interest \citep{2023MNRAS.524.4256M,2025ApJ...986..127N}. 
Therefore, we consider this assumption is also reasonable for FDM.

\subsection{Zoom-in simulations} \label{sec3.3}
We use high-resolution dark matter-only zoom-in simulations from \citet{2015ApJ...810...21M}, which are the same as those adopted in \citet{2020ApJ...893...48N}, to generate subhalo populations. 
The simulation suite contains 45 MW-mass host halos with virial masses between $1.18$ and $1.57 \times 10^{12}~M_\odot$. 
Subhalos in these simulations are well resolved down to a present-day maximum circular velocity of $V_{\max} \approx 9~\mathrm{km\,s^{-1}}$. 
In addition, we require each subhalo to have a peak maximum circular velocity $V_{\mathrm{peak}} > 10~\mathrm{km\,s^{-1}}$.

Because the detailed merger history of the MW, particularly the early accretion of a galaxy with the Large Magellanic Cloud (LMC) mass, can affect its satellite population \citep{2020MNRAS.495..743B}, we follow \citet{2020ApJ...893...48N} and adopt two host halos with assembly histories similar to that of the MW, which have virial masses of $1.26$ and $1.57 \times 10^{12}~M_\odot$, respectively.

To avoid potential systematic uncertainties introduced by using different simulations, we use the same high-resolution dark matter-only zoom-in simulations from \citet{2015ApJ...810...21M} for M31, even though its virial mass and merger history remain highly uncertain \citep{2018ApJ...864...34S,2019ApJ...872...24V,2023ApJ...948..104P,2024MNRAS.528.2653Z,2025arXiv250206948D}. 
Specifically, we select two host halos with virial masses of $1.19$ and $1.35 \times 10^{12}~M_\odot$ to account for the uncertainty in M31 assembly history. 
We have verified that our results are insensitive to this choice, as using different simulated host halos yields similar constraints. 
We discuss our treatment of the large uncertainty in M31 virial mass in Section~\ref{sec3.4}.

\subsection{Model parameters and statistical inference} \label{sec3.4}
The empirical galaxy--halo connection model includes eight parameters \citep{2020ApJ...893...48N}. 
The first two parameters describe the mapping between the subhalo peak maximum circular velocity $V_{\mathrm{peak}}$ and the satellite absolute magnitude $M_V$: the power-law slope of the luminosity function, $\alpha$, and the luminosity scatter, $\sigma_M$. 
The next two parameters characterize the galaxy formation efficiency based on the subhalo peak virial mass $M$: the peak virial mass at which the galaxy occupation fraction is 50\%, denoted as $\mathcal{M}_{50}$, and the parameter that determines the tilt of the galaxy occupation fraction, $\mathcal{S}_{\mathrm{gal}}$.
The fifth parameter, $\mathcal{B}$, quantifies the strength of subhalo disruption due to baryonic effects. 
The remaining three parameters describe the mapping between the subhalo virial radius at accretion and the satellite half-light radius $r_{1/2}$: the amplitude of the galaxy--halo size relation, $\mathcal{A}$, the scatter in half-light radius at fixed halo size, $\sigma_{\log_{10} R}$, and the power-law index of the size relation, $n$.
We also introduce the parameter $M_{\mathrm{hm}}$ to model the effects of FDM or WDM \citep{2021PhRvL.126i1101N}. 
Its impact enters through the subhalo mass function suppression described by Equations~(\ref{fdmshmf}) or (\ref{wdmshmf}), 
and $M_{\mathrm{hm}}$ can be converted into the corresponding dark matter particle mass using Equations~(\ref{fdmmhm}) or (\ref{wdmmhm}).

Subhalo abundance is known to scale linearly with host halo mass \citep{2012MNRAS.424.2715W,2015ApJ...810...21M,2025arXiv250912313T}. 
In addressing the uncertainty in host halo mass, \citet{2020ApJ...893...48N} simply scale the final results based on the average mass of the two realistic MW-like simulations and the $2\,\sigma$ upper limit of the MW mass predicted in \citet{2019MNRAS.484.5453C}, yielding a very conservative result.
In our work, we introduce the mass parameter $M_{\mathrm{host}}$, which represents the virial mass of the MW or M31. 
This parameter linearly scales $\mathcal{M}_{50}$ and $M_{\mathrm{hm}}$ relative to the simulated host halo mass $M_{\mathrm{sim}}$, such that $\mathcal{M}_{50} = \mathcal{M}_{50}^{\mathrm{(sim)}} (M_{\mathrm{host}}/M_{\mathrm{sim}})$ and 
$M_{\mathrm{hm}} = M_{\mathrm{hm}}^{\mathrm{(sim)}} (M_{\mathrm{host}}/M_{\mathrm{sim}})$.
We assign a log-normal prior within a $2\,\sigma$ range to the $M_{\mathrm{host}}$ parameter, so its role is to linearly scale $\mathcal{M}_{50}$ and $M_{\mathrm{hm}}$ using the mass prior information, effectively accounting for the uncertainties in the MW and M31 masses. 

For the MW, we use the results from \citet{2019MNRAS.484.5453C}, corresponding to $\log_{10}(M_{\mathrm{MW}})$ in the range of [12.00, 12.26]. 
For M31, we adopt the baseline prior from \citet{2024MNRAS.528.2653Z}, corresponding to $\log_{10}(M_{\mathrm{M31}})$ in the range of [11.81, 12.45]. 
At the same time, we note that M31's mass uncertainty is significant. 
The result from \citet{2023ApJ...948..104P} ($\sim 3.02 \times 10^{12}~M_\odot$) is about a factor of 2-3 higher than our baseline setting from \citet{2024MNRAS.528.2653Z} ($\sim 1.33 \times 10^{12}~M_\odot$). 
Therefore, we also consider the result from \citet{2023ApJ...948..104P}, corresponding to $\log_{10}(M_{\mathrm{M31}})$ in the range of [12.22, 12.74].
Note that while the mass definitions in \citet{2019MNRAS.484.5453C} and \citet{2024MNRAS.528.2653Z} differ from the one adopted in this work, we have consistently converted these values to ensure a uniform mass definition across our analysis.
The priors for the other parameters are consistent with those in \citet{2025ApJ...986..127N}.

To account for uncertainty in the observer's location, we fix the observer's distance to $8~\mathrm{kpc}$ for the MW \citep{2020ApJ...893...48N} and to $776~\mathrm{kpc}$ for M31 \citep{2022ApJ...938..101S}, and then draw a random viewing position on the sphere at the corresponding radius. 
For MW realizations, we additionally rotate each simulated host halo, so that the LMC analog lies at the observed sky position of the LMC. 
For M31 realizations, we rotate the simulated system, so that the host center projects to the observed sky position of M31.

Therefore, for the MW or M31, given these 10 parameters $\boldsymbol{\theta}$ and simulation data and using the binning method from \citet{2025ApJ...986..127N}, the number of predicted observable satellites in luminosity and surface brightness bin $i$ in a given realization is
\begin{align}
N_{\mathrm{pred},i} &= \sum_{s_i} p_{\mathrm{detect},s_i} \times \left( 1 - p_{\mathrm{disrupt},s_i} \right)\nonumber \\
&\times f_{\mathrm{gal},s_i} \times f_{\mathrm{non\text{-}CDM},s_i},
\label{numbin}
\end{align}
where $ s_i $ represents the mock satellites in bin $i$, $p_{\mathrm{detect},s_i} $ denotes the detection probability, which is determined by the observational selection functions, $ p_{\mathrm{disrupt},s_i} $ refers to the probability of disruption due to baryonic effects, $f_{\mathrm{gal},s_i} $ is the galaxy formation efficiency, and $f_{\mathrm{non\text{-}CDM},s_i} $ represents the subhalo mass function suppression.

The probability of observing $N_{\mathrm{obs},i}$ satellites in bin $i$ is \citep{2019ApJ...873...34N,2020ApJ...893...48N}
\begin{align}
P(&N_{\mathrm{obs},i}|\boldsymbol{\theta}) = 
\Big(\frac{N_{\mathrm{real}} + 1}{N_{\mathrm{real}}}\Big)^{-({N}_{i} + 1)} \nonumber\\
&\times (N_{\mathrm{real}} + 1)^{-N_{\mathrm{obs},i}}
\frac{\Gamma({N}_{i} + N_{\mathrm{obs},i} + 1)}{\Gamma(N_{\mathrm{obs},i} + 1)\Gamma({N}_{i} + 1)},
\label{poisson_like}
\end{align}
where $N_{\mathrm{real}}$ is the number of model realizations given $\boldsymbol{\theta}$ and ${N}_{i}\equiv \sum_{j=1}^{{N_\mathrm{real}}}N_{\mathrm{pred},ij}$.
For each simulated host halo, we generate 30 mock realizations corresponding to different absolute magnitudes, half-light radii, positions, and galaxy formation realizations. 
Because we use two simulated host halos for both the MW and M31 to account for uncertainties in their halo properties, this yields a total of $N_{\mathrm{real}} = 60$ mock realizations.
We have verified that 60 realizations are sufficient for convergence.

Before constructing the total likelihood function, we clarify that when jointly modeling the MW and M31 satellite populations, we treat $\mathcal{M}_{50}$, $\mathcal{S}_{\mathrm{gal}}$, and $M_{\mathrm{hm}}$ as shared parameters, while all other parameters are allowed to vary independently for the MW and M31.
We make this choice because the mass of M31 is highly uncertain and may exceed the mass range covered by our simulation suite \citep{2015ApJ...810...21M,2023ApJ...948..104P,2024MNRAS.528.2653Z}. 
Although our galaxy--halo connection model is sufficiently flexible to reproduce the observed data even when the host mass lies outside the simulated range, the inferred parameters in such cases would be systematically biased (see Appendix~\ref{app} for a detailed discussion). 
In contrast, $\mathcal{M}_{50}$, $\mathcal{S}_{\mathrm{gal}}$, and $M_{\mathrm{hm}}$ can be reliably scaled according to the mass priors, yielding physically consistent values.

We denote the MW- and M31-specific parameter sets by $\boldsymbol{\theta}_{\mathrm{MW}}$ and $\boldsymbol{\theta}_{\mathrm{M31}}$ and the shared parameters by $\boldsymbol{\theta}_{\mathrm{share}}$. 
Given the vectors of observed satellite counts from DES, PS1, and PAndAS, $\boldsymbol{N}_{\mathrm{DES}}=\{N_{\mathrm{obs},i}^{\mathrm{DES}}\}$, $\boldsymbol{N}_{\mathrm{PS1}}=\{N_{\mathrm{obs},i}^{\mathrm{PS1}}\}$, and $\boldsymbol{N}_{\mathrm{PA}}=\{N_{\mathrm{obs},i}^{\mathrm{PA}}\}$, the total likelihood is
\begin{align}
&P(\boldsymbol{N}_{\mathrm{DES}},\boldsymbol{N}_{\mathrm{PS1}},\boldsymbol{N}_{\mathrm{PA}}\mid \boldsymbol{\theta}_{\mathrm{MW}},\boldsymbol{\theta}_{\mathrm{M31}},\boldsymbol{\theta}_{\mathrm{share}})
= 
\nonumber
\\  &\prod_{i \in \mathrm{DES}} P\!\left(N_{\mathrm{obs},i}^{\mathrm{DES}} \mid \boldsymbol{\theta}_{\mathrm{MW}}, \boldsymbol{\theta}_{\mathrm{share}}\right) \prod_{i \in \mathrm{PS1}} P\!\left(N_{\mathrm{obs},i}^{\mathrm{PS1}} \mid \boldsymbol{\theta}_{\mathrm{MW}}, \boldsymbol{\theta}_{\mathrm{share}}\right) \nonumber
\\  &\times \prod_{i \in \mathrm{PA}} P\!\left(N_{\mathrm{obs},i}^{\mathrm{PA}} \mid \boldsymbol{\theta}_{\mathrm{M31}}, \boldsymbol{\theta}_{\mathrm{share}}\right),
\label{likelihood}
\end{align}
where $P(\cdot\,|\,\cdot)$ is the per-bin likelihood in Equation (\ref{poisson_like}) with the expected counts given by Equation (\ref{numbin}).
The joint posterior over all parameters is
\begin{align}
&P\!\left(\boldsymbol{\theta}_{\mathrm{MW}},\boldsymbol{\theta}_{\mathrm{M31}},\boldsymbol{\theta}_{\mathrm{share}}
\mid \boldsymbol{N}_{\mathrm{DES}},\boldsymbol{N}_{\mathrm{PS1}},\boldsymbol{N}_{\mathrm{PA}}\right) \nonumber
 \\ &\propto P(\boldsymbol{N}_{\mathrm{DES}},\boldsymbol{N}_{\mathrm{PS1}},\boldsymbol{N}_{\mathrm{PA}}\mid \boldsymbol{\theta}_{\mathrm{MW}},\boldsymbol{\theta}_{\mathrm{M31}},\boldsymbol{\theta}_{\mathrm{share}}) 
 \nonumber
 \\ &\times \,\pi_{\mathrm{MW}}\!\left(\boldsymbol{\theta}_{\mathrm{MW}}\right)\,
\pi_{\mathrm{M31}}\!\left(\boldsymbol{\theta}_{\mathrm{M31}}\right)\,
\pi_{\mathrm{share}}\!\left(\boldsymbol{\theta}_{\mathrm{share}}\right),
\label{tpost}
\end{align}
where $\pi_{\mathrm{MW}}(\boldsymbol{\theta}_{\mathrm{MW}})$ and $\pi_{\mathrm{M31}}(\boldsymbol{\theta}_{\mathrm{M31}})$ are the priors for the MW- and M31-specific parameters, and $\pi_{\mathrm{share}}(\boldsymbol{\theta}_{\mathrm{share}})$ is the prior for the shared parameters. 
The normalization constant (evidence) is omitted as it is independent of the parameters.
The marginal posterior for the shared parameters is:
\begin{align}
\label{posterior_both}
P&\!\left(\boldsymbol{\theta}_{\mathrm{share}}
\mid \boldsymbol{N}_{\mathrm{DES}}, \boldsymbol{N}_{\mathrm{PS1}},\boldsymbol{N}_{\mathrm{PA}}\right) =
\int \mathrm{d}\boldsymbol{\theta}_{\mathrm{MW}}\,
\int \mathrm{d}\boldsymbol{\theta}_{\mathrm{M31}}\,
\nonumber
\\ & \quad \
P\!\left(\boldsymbol{\theta}_{\mathrm{MW}},\boldsymbol{\theta}_{\mathrm{M31}},\boldsymbol{\theta}_{\mathrm{share}}
\mid \boldsymbol{N}_{\mathrm{DES}},\boldsymbol{N}_{\mathrm{PS1}},\boldsymbol{N}_{\mathrm{PA}}\right).
\end{align}
We use \texttt{emcee} \citep{2013PASP..125..306F} in the constraint process for the model parameters. 
Specifically, we run 112 walkers with $10^{5}$ steps for each walker, and discard the first 10 maximum autocorrelation lengths as burn-in ( $\sim10^{4}$ steps for each walker) to sample the posterior distributions or the probability distribution functions (PDFs) of the parameters.

\begin{figure}
	\centering
	\begin{minipage}{\linewidth}
		\centering
		\includegraphics[width=\linewidth]{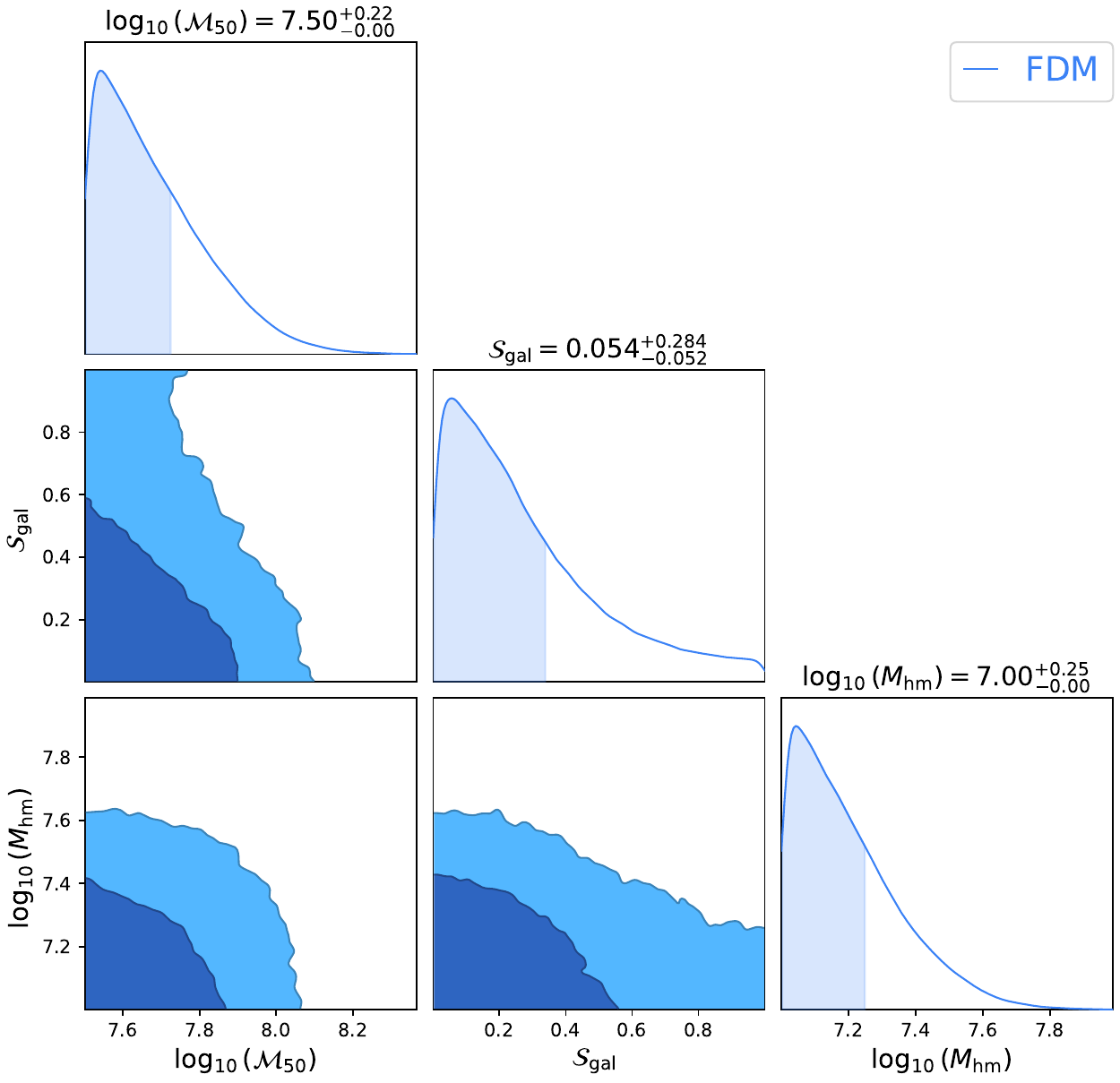}	
	\end{minipage}
    \caption{The PDF and contour maps (68\% and 95\% CLs) of the shared parameters in the FDM model. The shaded regions in the one-dimensional PDFs and the quoted parameter values correspond to 68\% CL.  The parameters $\mathcal{M}_{50}$ and $M_{\mathrm{hm}}$ are in units of $M_\odot$, and $\mathcal{S}_{\mathrm{gal}}$ is in dex.}
    \label{Joint_sp_m_single_fdm.pdf}
\end{figure}

\begin{figure}
	\centering
	\begin{minipage}{\linewidth}
		\centering
		\includegraphics[width=\linewidth]{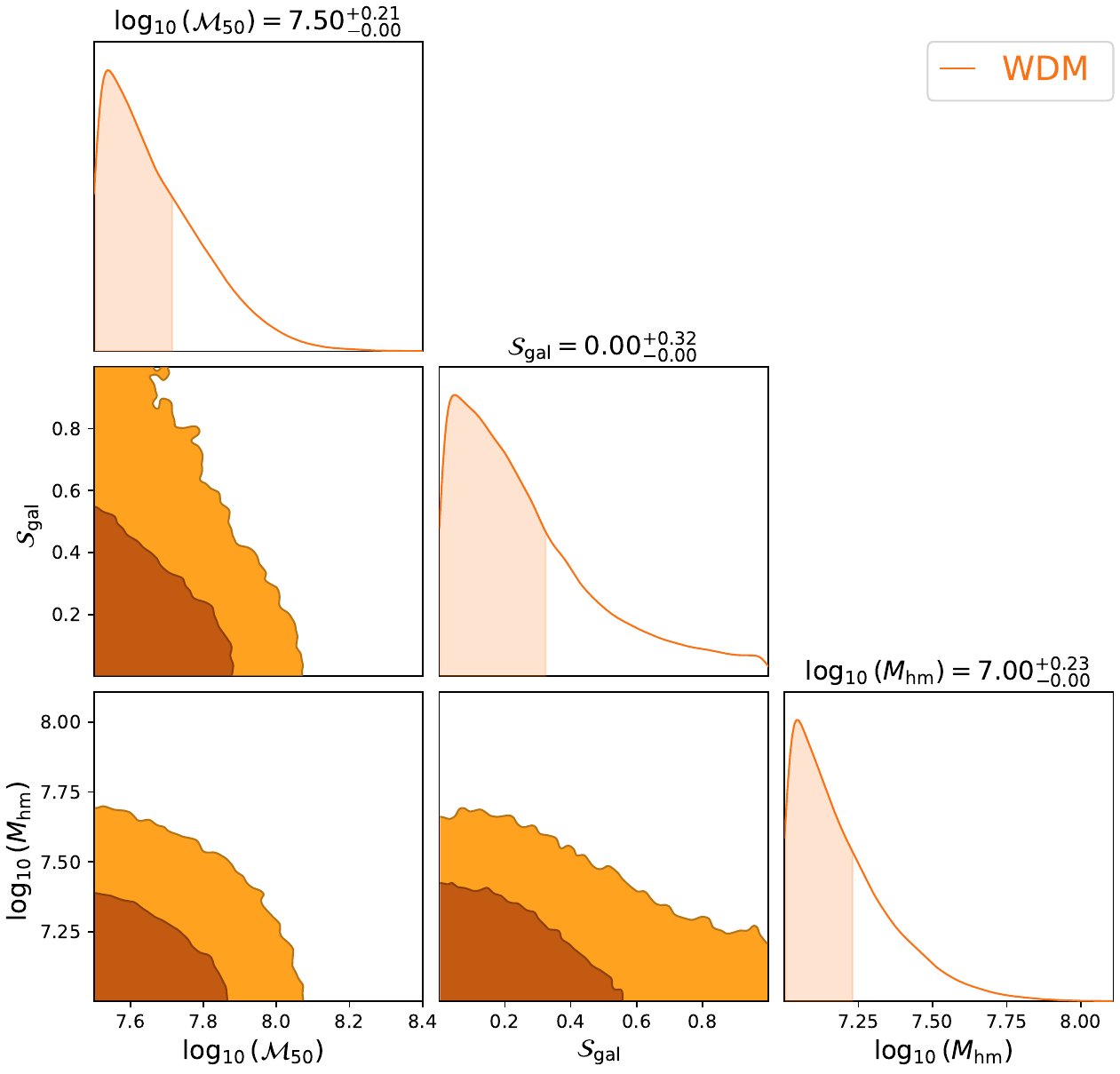}	
	\end{minipage}
    \caption{Same as Figure \ref{Joint_sp_m_single_fdm.pdf}, but for the WDM case.}
    \label{Joint_sp_m_single_wdm.pdf}
\end{figure}

\section{Constraint results} \label{sec4}

\subsection{Fuzzy dark matter} \label{sec4.1}
In Figure~\ref{Joint_sp_m_single_fdm.pdf}, we show the marginalized PDFs and contour maps for the shared parameters in the FDM model.
From the PDFs, we derive $M_{\mathrm{hm}} < 3.12 \times 10^{7}~M_\odot$ at 95\% confidence level (CL), which can be converted into a lower limit on the FDM particle mass using Equation~(\ref{fdmmhm}), yielding
\begin{equation}
m_{\mathrm{FDM}} > 1.74 \times 10^{-20}~\mathrm{eV}\ (95\% \ \mathrm{CL}).
\label{fdmcon}
\end{equation}
This constraint improves upon the result obtained using only the MW satellite population by \citet{2025ApJ...986..127N}, $m_{\mathrm{FDM}} > 1.4 \times 10^{-20}~\mathrm{eV}$, by about 24\%.
The precision of our limit is also comparable to that from the Lyman-$\alpha$ forest analysis in \citet{2021PhRvL.126g1302R}, which yields $m_{\mathrm{FDM}} > 2.0 \times 10^{-20}~\mathrm{eV}$.

Besides, to obtain a more prior-independent constraint, we also adopt a posterior ratio of 20:1, corresponding to the point where the posterior probability drops to one-twentieth of its maximum value \citep{2023MNRAS.524L..84P,2024MNRAS.535.1652K,2025PhRvD.111f3079T}. 
Our lower prior bound on $M_{\mathrm{hm}}$, $10^{7}~M_\odot$, coincides with the region where the posterior distribution flattens out, and we have verified the robustness of the 20:1 posterior ratio result by varying the prior range. 
For the FDM case, this criterion yields $M_{\mathrm{hm}} < 4.17 \times 10^{7}~M_\odot$, corresponding to
\begin{equation}
m_{\mathrm{FDM}} > 1.42 \times 10^{-20}~\mathrm{eV}\ (\text{20:1}\ \mathrm{posterior\ ratio}).
\label{fdmcon20}
\end{equation}
Our result is about three times stronger than the posterior odds ratio of 20:1 constraint derived from strong gravitational lensing by \citet{2023MNRAS.524L..84P}, who obtained $m_{\mathrm{FDM}} > 4.4 \times 10^{-21}~\mathrm{eV}$.
We note that \citet{2023MNRAS.524L..84P} and \citet{2025PhRvD.111f3079T} adopt a posterior odds ratio, while \citet{2021ApJ...917....7N} use a likelihood ratio. 
However, because our posterior flattens toward the lower prior boundary and we adopt a uniform prior in $\log_{10} (M_{\mathrm{hm}})$, these criteria reduce to the same effective threshold as our posterior ratio.

We present the full PDFs and contour maps of the MW and M31 parameters for the FDM case in Appendix~\ref{app}. 
We have verified that, aside from the shared parameters, the posterior distributions of the other parameters are consistent with those obtained from the separate MW and M31 fits. 
Furthermore, we rerun our model using the larger M31 mass prior from \citet{2023ApJ...948..104P}, and obtain similar and slightly weaker constraint results with $m_{\mathrm{FDM}} > 1.68 \times 10^{-20}\,\mathrm{eV}$ (95\% CL) and $m_{\mathrm{FDM}} > 1.34 \times 10^{-20}\,\mathrm{eV}$ (20:1 posterior ratio).

\subsection{Warm dark matter} \label{sec4.2}
Figure~\ref{Joint_sp_m_single_wdm.pdf} shows the marginalized PDFs and contour maps of the shared parameters in the WDM model. 
The marginalized posterior yields $M_{\mathrm{hm}} < 3.23 \times 10^{7}~M_\odot$ at 95\% CL, corresponding to a lower limit on the WDM particle mass using Equation~(\ref{wdmmhm}), and we have
\begin{equation}
m_{\mathrm{WDM}} > 6.20~\mathrm{keV}\ (95\% \ \mathrm{CL}).
\label{wdmcon}
\end{equation}
This constraint represents a $\sim$5\% improvement over the result obtained using only the MW satellite population by \citet{2025ApJ...986..127N} with $m_{\mathrm{WDM}} > 5.9~\mathrm{keV}$.
At a posterior ratio of 20:1, we obtain $M_{\mathrm{hm}} < 4.23 \times 10^{7}~M_\odot$, corresponding to
\begin{equation}
m_{\mathrm{WDM}} > 5.75~\mathrm{keV}\ (\text{20:1}\ \mathrm{posterior\ ratio}).
\label{wdmcon20}
\end{equation}

Since the posterior distributions for all parameters in the WDM case are very similar to those in the FDM case, we do not provide figures showing the posterior distributions of all the MW and M31 parameters for the WDM case.
We also rerun our model using the larger M31 mass prior from \citet{2023ApJ...948..104P}, obtaining $m_{\mathrm{WDM}} > 6.07~\mathrm{keV}$ (95\% CL) and $m_{\mathrm{WDM}} > 5.61~\mathrm{keV}$ (20:1 posterior ratio).

We note that in \citet{2025ApJ...986..127N}, the inferred $M_{\mathrm{hm}}$ for WDM is slightly lower than that for FDM, whereas our results show the opposite trend. 
This difference also explains why our improvement in the WDM constraint is only marginal. 
A possible reason is that our number of realizations, $N_{\mathrm{real}}$, is much larger than that used in \citet{2025ApJ...986..127N}, making our model more sensitive to subtle differences in the subhalo population. 
As shown in Figure~\ref{transfer_shmf_relative.pdf}, at the most sensitive mass scale around $10^{8}\,M_\odot$, the subhalo mass function for WDM is slightly higher than that for FDM at the same $M_{\mathrm{hm}}$.
To compensate for this difference, the inferred $M_{\mathrm{hm}}$ for WDM tends to be slightly larger, resulting in the extended tail as shown in Figure~\ref{Joint_sp_m_single_wdm.pdf}.

Compared to \citet{2025ApJ...986..127N}, our improvements arise mainly from two aspects: first, we adopt a more flexible mass-scaling approach; second, we include the M31 satellite population in our joint analysis.  
Since the constraints are primarily sensitive to the faintest satellites, incorporating the M31 satellite population strengthens the limits only at the few-percent level, which we summarize as an overall improvement of $\sim 10\%$ relative to using the MW sample alone. 
This also explains why the results are insensitive to the methodological differences in selection functions and the inclusion or exclusion of the $M_{V} \simeq -6~\mathrm{mag}$ M31 satellite, since the M31 satellite population does not dominate the final constraints.

Moreover, because the mass and accretion history of M31 are highly uncertain, we do not treat all parameters as shared parameters, as was done in \citet{2024ApJ...967...61N}, when presenting our final constraints.
We have verified that treating all parameters as shared would further tighten the constraints by about 5\% relative to our adopted approach. 
However, as discussed above, enforcing the sharing of all parameters could introduce unphysical biases given the significant uncertainties in M31's mass and accretion history. 
Therefore, this $\sim 5\%$ improvement likely stems from the compensation for these biased parameters rather than a genuine physical signal.

\begin{figure*}
	\centering
	\begin{minipage}{\linewidth}
		\centering
		\includegraphics[width=\linewidth]{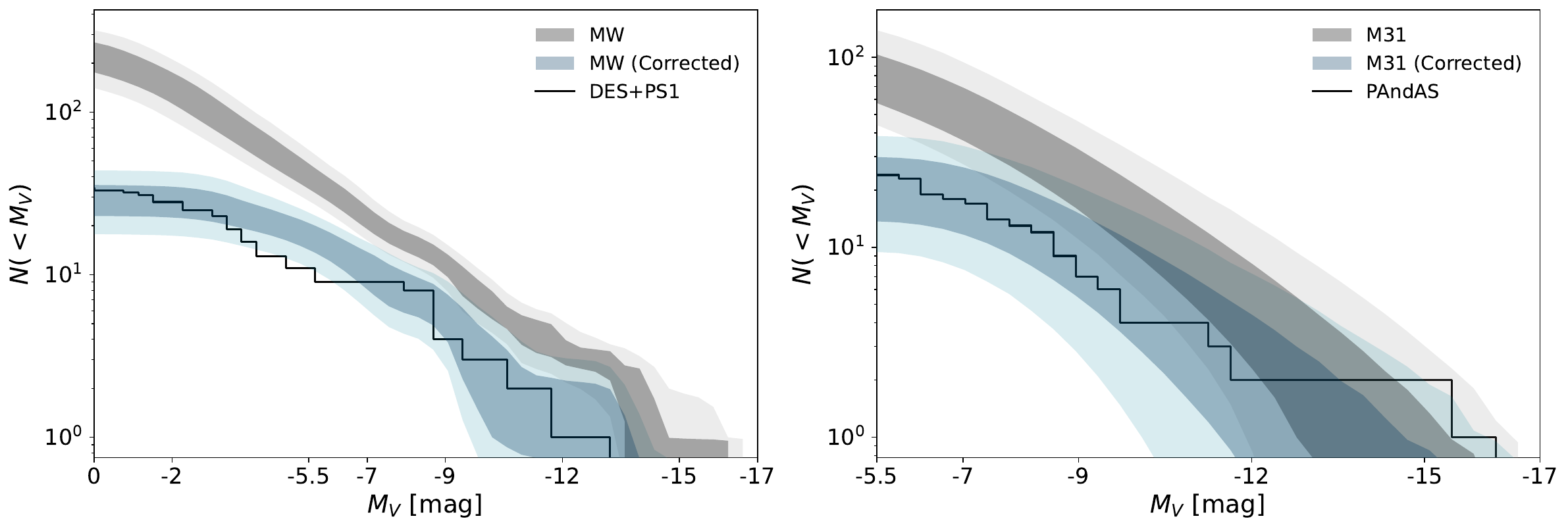}	
	\end{minipage}
    \caption{Luminosity functions of satellites located within $300~\mathrm{kpc}$ from the centers of the MW (left panel) and M31 (right panel), derived from the FDM results in our modeling framework. The gray shaded regions represent the predicted luminosity functions, while the blue shaded regions show the predicted observable luminosity functions after applying the observational selection functions. The black stepped lines indicate the observed satellite counts from DES and PS1 (left) and PAndAS (right). Dark and light shaded regions correspond to the 68\% and 95\% CLs, respectively.}
    \label{satellite_luminosity_function.pdf}
\end{figure*}

\section{Additional Results and Discussion} \label{sec5}

\subsection{Luminosity functions of the MW and M31} \label{sec5.1}
Using the posterior distributions inferred from our modeling framework, and adopting the FDM results as a representative case, we predict the satellite luminosity functions for both the MW and M31.
As shown in Figure~\ref{satellite_luminosity_function.pdf}, our model reproduces the observed luminosity functions of both the MW and M31 remarkably well, with the observational data falling within the 68\% CLs. 
This validates the reliability of our modeling framework and the resulting constraints. 
The fact that the M31 luminosity function is well-reproduced further suggests that the significant uncertainties in M31's halo mass and accretion history do not substantially bias our results. 
In other words, given the current observational precision, our constraints are relatively insensitive to the specific details of M31's host halo properties. 
Furthermore, when performing the same check for a model where all parameters are shared between the MW and M31, we find that the MW luminosity function is slightly overestimated, while that of M31 is significantly underestimated. 
This discrepancy reinforces the conclusion that enforcing a fully shared parameter space yields unphysical results.

For satellites with $M_V < 0~\mathrm{mag}$, we predict a total of $222^{+47}_{-46}$ satellites within $300~\mathrm{kpc}$ of the MW, which is consistent with the result of \citet{2020ApJ...893...48N}. 
For satellites with $M_V < -5.5~\mathrm{mag}$, we predict $41^{+7}_{-7}$ and $80^{+24}_{-23}$ satellites within $300~\mathrm{kpc}$ of the MW and M31, respectively.
These values are in good agreement with the results from \citet{2025arXiv250912313T} and \citet{2023ApJ...952...72D}. 
We find that the abundance of satellites in M31 is approximately twice that of the MW, a trend that was also noted by \citet{2025arXiv250912313T}. 
Considering that satellite abundance is known to scale roughly linearly with host halo mass, this may suggest that the halo mass of M31 is approximately twice that of the MW.
However, the enhanced satellite abundance around M31 may also be partly explained by its past interaction with M33 \citep{2013MNRAS.430...37C,2025arXiv250206948D}.

\subsection{Anisotropy in the M31 satellite distribution} \label{sec5.2}
As more M31 satellites have been discovered, it has become evident that their spatial distribution may be not isotropic. 
Specifically, among the 24 M31 satellites used in this work, only three are located on the side opposite the MW \citep{2022ApJ...938..101S,2023ApJ...952...72D}. 
To examine whether this anisotropy could arise from observational selection effects, we use the posterior distributions inferred from our modeling framework to simulate $10^{5}$ realizations, each representing an observation of 24 M31 satellites. 

\begin{figure}
	\centering
	\begin{minipage}{\linewidth}
		\centering
		\includegraphics[width=\linewidth]{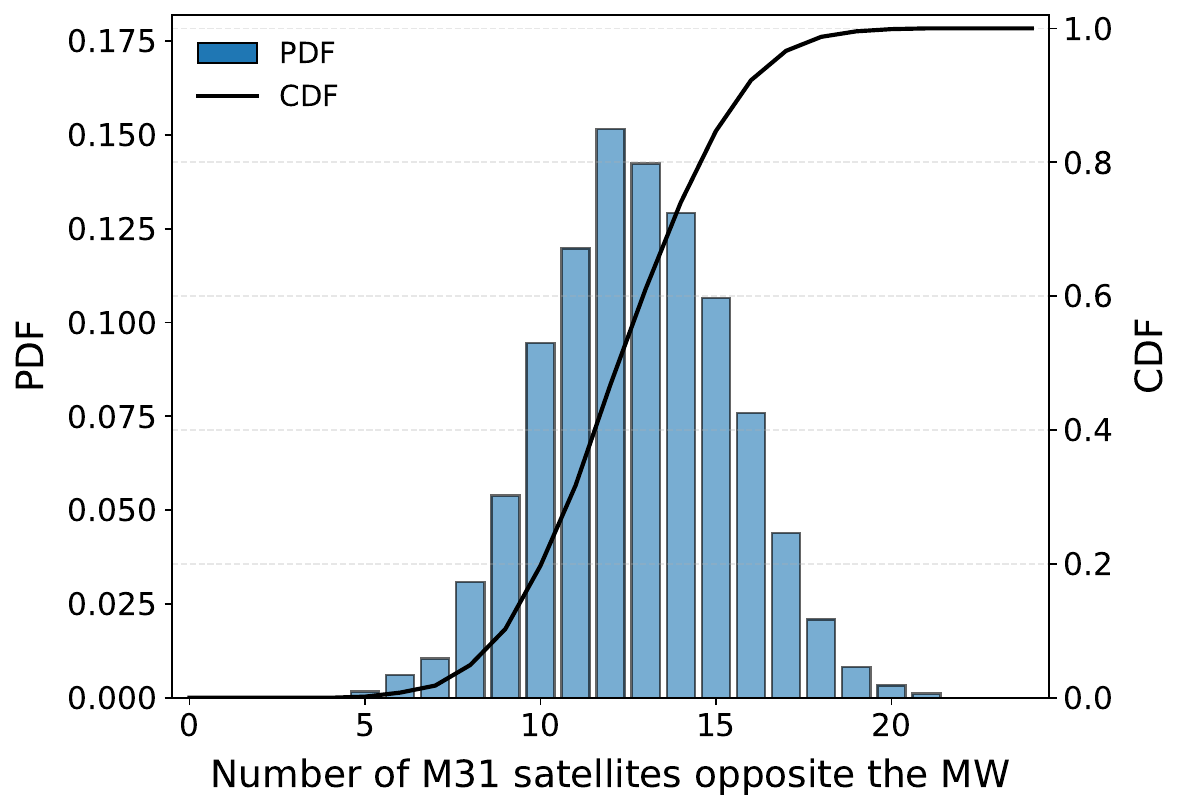}	
	\end{minipage}
    \caption{The PDF (blue histogram) and CDF (black curve) of the number of M31 satellites located on the side opposite the MW, based on $10^{5}$ simulated realizations conditioned on the 24 observed satellites, using the FDM results from our modeling framework.}
    \label{probability_density_function.pdf}
\end{figure}

We then obtain the PDF and cumulative distribution function (CDF) of the number of satellites located opposite the MW (see Figure~\ref{probability_density_function.pdf}). 
In all $10^{5}$ simulated realizations, none produce as few as three satellites on the far side. 
Even the probability of having eight satellites on the far side corresponds to a CDF value below $0.05$. 
This indicates that, even after accounting for distance uncertainties, the observed distribution cannot be explained by selection effects alone, consistent with the conclusions of previous studies \citep{2022ApJ...938..101S,2023ApJ...952...72D,2025NatAs...9..692K}. 
Interestingly, although detectability decreases with increasing distance, the larger observable volume counteracts this decline. 
As a result, if the M31 satellite distribution is isotropic, the probability of detecting more satellites on the far side than on the near side would in fact be slightly greater than 50\%.
Finally, we note that the strong anisotropy of the M31 satellites may, in principle, affect abundance-based constraints, but no reliable framework currently exists to incorporate this effect. 
In our analysis, we do not include any spatial information in our binning scheme. 
We have also verified that using different simulated host halos yields consistent constraints within our modeling framework. 
The fact that our model is in good agreement with the observed data (as shown in Figure~\ref{satellite_luminosity_function.pdf}) further reinforces the reliability of our current approach.
We will explore this issue in more detail in future work.

\section{Summary and Outlook} \label{sec6}
In this work, we develop a modeling framework and perform a joint analysis of the MW and M31 satellite systems to  study the properties of both FDM and WDM. We use the empirical galaxy--halo connection model described in \citet{2020ApJ...893...48N} together with the updated transfer functions and subhalo mass functions to construct the framework, and adopt the observational data of MW and M31 satellite galaxies from DES, PS1, and PAndAS, including their corresponding selection functions.
We further account for the uncertain virial masses of the MW and M31 by introducing priors on their host halo masses and linearly scaling the relevant model parameters. 

We find that, for the FDM case, $m_{\mathrm{FDM}} > 1.74 \times 10^{-20}~\mathrm{eV}$ (95\% CL) and $m_{\mathrm{FDM}} > 1.42 \times 10^{-20}~\mathrm{eV}$ (20:1 posterior ratio). 
For thermal-relic WDM, we find $m_{\mathrm{WDM}} > 6.20~\mathrm{keV}$ (95\% CL) and $m_{\mathrm{WDM}} > 5.75~\mathrm{keV}$ (20:1 posterior ratio). 
These results represent a moderate improvement over the MW-only constraints given by the previous works. 
Our analysis therefore provides the strongest constraints to date on FDM and WDM models derived from satellite galaxy populations in the Local Group. We also verify that adopting extreme assumptions for the mass of M31 weakens our constraints only slightly.
These limits can be combined with complementary probes such as strong gravitational lensing and the Lyman-$\alpha$ forest to achieve even tighter bounds \citep{2021ApJ...917....7N,2021MNRAS.506.5848E}.

Looking ahead, next-generation surveys such as the Rubin Observatory Legacy Survey of Space and Time (LSST; \citealt{2025OJAp....8E..89T}), Chinese Space Station Survey Telescope (CSST; \citealt{2023MNRAS.523..876Q,2025arXiv250704618C}), and Dark Energy Spectroscopic Instrument (DESI; \citealt{2023ApJ...944....1D}) will greatly expand the known satellite populations in the Local Group, including systems fainter than those currently detected. 
In addition, these surveys will also improve measurements of the virial masses and accretion histories of the MW and M31. 
Combined with advances in high-precision, large-volume dark-matter simulations that probe a broader range of host masses and assembly histories, as well as continued progress in observational selection functions, future analyses will be able to achieve more accurate constraints on the properties of dark matter.

%% Please use the acknowledgment and contribution environments. This will 
%% be anonomyized when the "anonymous" style option is used. 
\begin{acknowledgments}
We thank Ethan Nadler for the help with implementing the empirical galaxy--halo connection model  code and for his helpful suggestions. J.X.L. and Y.G. acknowledge the support from the CAS Project for Young Scientists in Basic Research (No. YSBR-092), and National Key R\&D Program of China grant Nos. 2022YFF0503404 and 2020SKA0110402. 
KL was supported by National Key R\&D Programme of China (no. 2024YFC2207400). This work is also supported by science research grants from the China Manned Space Project with grant Nos. CMS-CSST-2025-A02, CMS-CSST-2021-B01, and CMS-CSST-2021-A01.
\end{acknowledgments}

\software{\texttt{Astropy} \citep{astropy:2013, astropy:2018, astropy:2022},
\texttt{ChainConsumer} \citep{Hinton2016},
\texttt{emcee} \citep{2013PASP..125..306F},
\texttt{healpy} \citep{2005ApJ...622..759G, Zonca2019},
\texttt{Matplotlib} \citep{Hunter:2007},
\texttt{NumPy} \citep{ harris2020array},
\texttt{pandas} \citep{mckinney-proc-scipy-2010, reback2020pandas},
\texttt{SciPy} \citep{2020SciPy-NMeth}
          }

\appendix
\section{Posterior distributions of model parameters} \label{app}
\begin{figure*}[ht!]
	\centering
	\begin{minipage}{\linewidth}
		\centering
		\includegraphics[width=\linewidth]{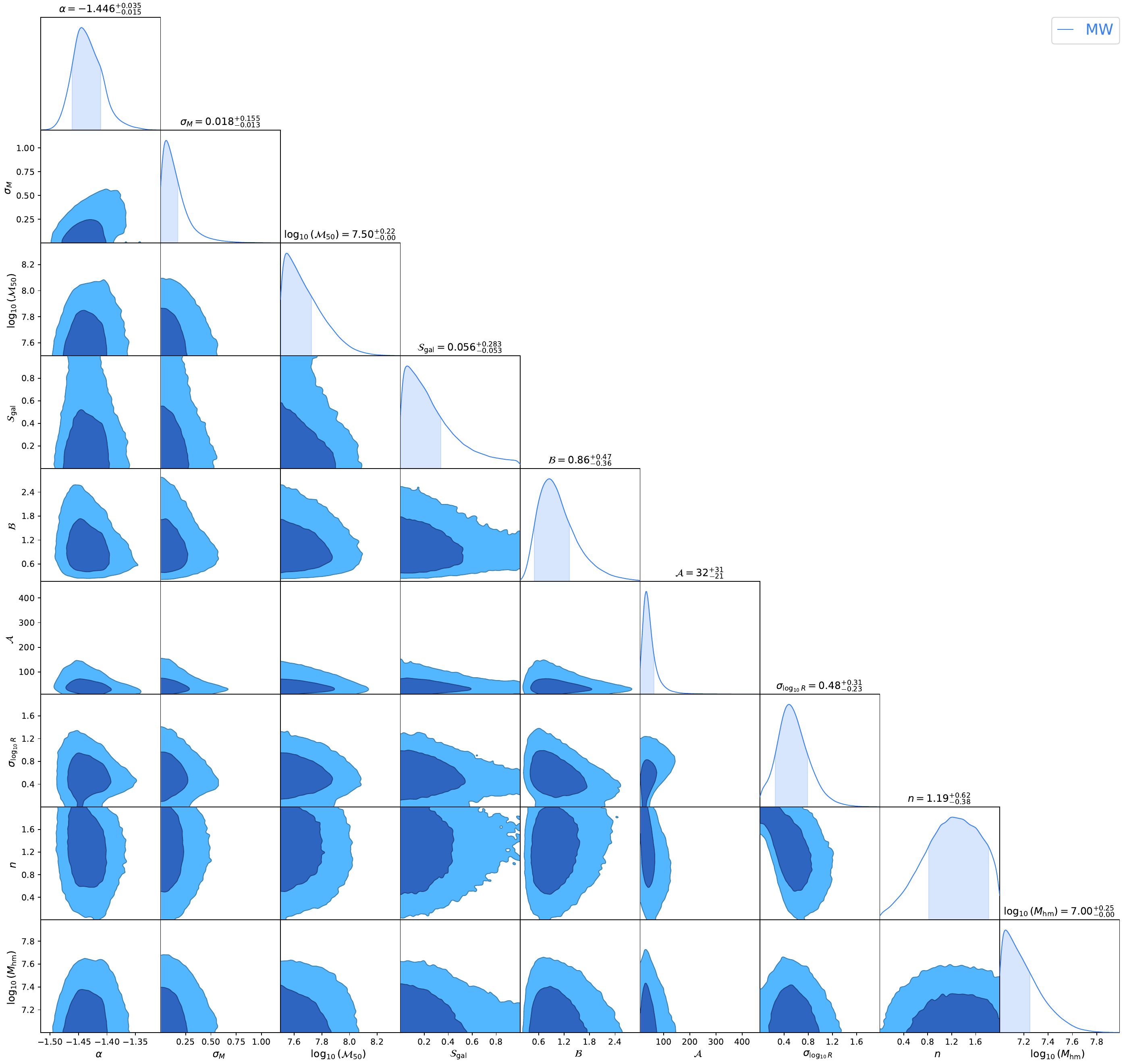}
	\end{minipage}
    \caption{The PDF and contour maps (68\% and 95\% CLs) of the MW parameters for the FDM case. The shaded regions in the one-dimensional marginalized PDFs and the parameter summaries denote 68\% CL. The parameters $\sigma_M$, $\mathcal{S}_{\mathrm{gal}}$, and $\sigma_{\log_{10} R}$ are in dex, $\mathcal{M}_{50}$ and $M_{\mathrm{hm}}$ are in units of $M_\odot$, $\mathcal{A}$ is reported in pc, and $\alpha$, $n$, and $\mathcal{B}$ are dimensionless.}
    \label{Joint_sp_mw_single.pdf}
\end{figure*}

\begin{figure*}[ht!]
	\centering
	\begin{minipage}{\linewidth}
		\centering
		\includegraphics[width=\linewidth]{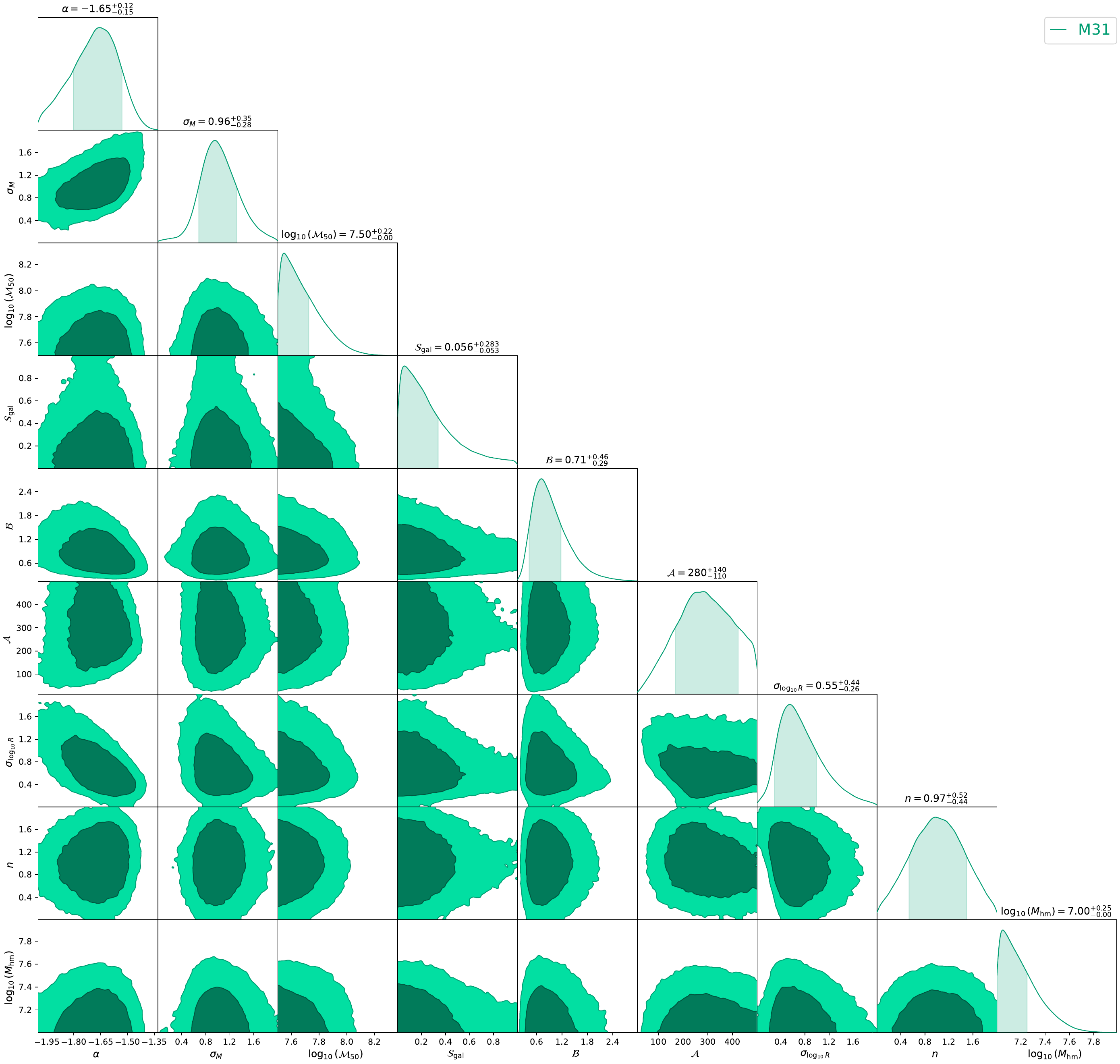}	
	\end{minipage}
    \caption{Same as Figure \ref{Joint_sp_mw_single.pdf}, but for the M31 case.}
    \label{Joint_sp_m31_single.pdf}
\end{figure*}

In Figures~\ref{Joint_sp_mw_single.pdf} and \ref{Joint_sp_m31_single.pdf}, we show the PDFs and contour maps (68\% and 95\% CLs) of the MW and M31 parameters for the FDM case, using the model introduced in Section~\ref{sec3}. 
We do not show the $M_{\mathrm{host}}$ parameter because its posterior closely follows the adopted prior.

As discussed in Section~\ref{sec3.4}, the posterior distributions of the M31-specific parameters inferred from our model are likely to be biased. 
This is because the virial mass of M31 is probably larger than the mass range of the simulated host halos we use, and its accretion history remains highly uncertain \citep{2018ApJ...864...34S,2019ApJ...872...24V,2023ApJ...948..104P,2024MNRAS.528.2653Z,2025arXiv250206948D}. 
As shown in Figures~\ref{Joint_sp_mw_single.pdf} and \ref{Joint_sp_m31_single.pdf}, the MW and M31 exhibit significant differences in three parameters: the power-law slope of the luminosity function, $\alpha$, the luminosity scatter, $\sigma_M$, and the amplitude of the galaxy--halo size relation, $\mathcal{A}$. 
This behavior is consistent with our expectations. 

However, the faintest satellite in our M31 sample has an absolute magnitude of $M_V \simeq -6.0~\mathrm{mag}$, while the faintest MW satellite reaches $M_V \simeq 0~\mathrm{mag}$. 
The absolute magnitude of $M_V \simeq -6.0~\mathrm{mag}$ lies at the luminosity scale that is most sensitive to the reionization. 
Therefore, the fainter systems, which in our model appear only among the MW satellites, tend to form most of their stars before reionization. 
As a result, they develop more compact stellar distributions and deviate from the general size-luminosity relation \citep{2022MNRAS.516.3944M,2025arXiv250912313T,2025arXiv250206948D}.
Their luminosity function may also deviate from a simple power law \citep{2018ApJ...863..123B,2022MNRAS.516.3944M}.
This effect could partially explain the differences in the inferred parameters between the MW and M31. 
To test this, we rerun our model using only MW satellites with $M_V < -6.0~\mathrm{mag}$. 
Although the parameter uncertainties increase due to the reduced data size, the discrepancy between the MW and M31 parameters decreases to some extent (with noticeable shifts in their maximum-posterior values), especially for the amplitude of the galaxy--halo size relation, $\mathcal{A}$.

\bibliography{sample701}{}
\bibliographystyle{aasjournalv7}

\end{document}